\documentstyle[preprint,aps,epsfig,epsf]{revtex} 
\def\npb#1{Nucl.\ Phys.\ B\, {\bf #1}}
\def\npbps#1{Nucl.\ Phys.\ (Proc. Suppl.) B\, {\bf #1}}
\def\plb#1{Phys.\ Lett.\ B\, {\bf #1}}

\def\prd#1{Phys.\ Rev.\ D\, {\bf #1}}
\def\prl#1{Phys.\ Rev.\ Lett.\ {\bf#1}}

\def\rmp#1{Rev.\ Mod.\ Phys.\ {\bf #1}}

\def\PHYSICA #1 #2 #3 {{\sl Physica}~{\bf#1} #2 (#3)}
\def\MPL #1 #2 #3 {{\sl Mod.~Phys.~Lett.}~{\bf#1} #2 (#3)}
\def\NPB #1 #2 #3 {{\sl Nucl.~Phys.}~{\bf #1} #2 (#3)}
\def\NPBPS #1 #2 #3 {{\sl Nucl.~Phys.~B~(Proc. Suppl.)}~{\bf #1} #2 (#3)}
\def\PLB #1 #2 #3 {{\sl Phys.~Lett.}~{\bf #1} #2 (#3)}
\def\PR #1 #2 #3 {{\sl Phys.~Rep.}~{\bf#1} #2 (#3)}
\def\PRD #1 #2 #3 {{\sl Phys.~Rev.}~{\bf #1} #2 (#3)}
\def\PRL #1 #2 #3 {{\sl Phys.~Rev.~Lett.}~{\bf#1} #2 (#3)}
\def\RMP #1 #2 #3 {{\sl Rev.~Mod.~Phys.}~{\bf#1} #2 (#3)}
\def\ZPC #1 #2 #3 {{\sl Z.~Phys.}~{\bf #1} #2 (#3)}
\def\IJMP #1 #2 #3 {{\sl Int.~J.~Mod.~Phys.}~{\bf#1} #2 (#3)}

\newcommand{\nc}{\newcommand}
\nc{\beq}{\begin{equation}}   \nc{\eeq}{\end{equation}}
\nc{\bea}{\begin{eqnarray}}   \nc{\eea}{\end{eqnarray}}
\nc{\baa}{\begin{array}}      \nc{\eaa}{\end{array}}
\nc{\bit}{\begin{itemize}}    \nc{\eit}{\end{itemize}}
\nc{\ben}{\begin{enumerate}}  \nc{\een}{\end{enumerate}}
\nc{\bce}{\begin{center}}     \nc{\ece}{\end{center}}
\def\beqa{\begin{eqnarray}}
\def\eeqa{\end{eqnarray}}
\def\be{\beq}
\def\ee{\eeq}
\def\to{\rightarrow}

\def\lsim{\mathrel{\raise.3ex\hbox{$<$\kern-.75em\lower1ex\hbox{$\sim$}}}}
\def\gsim{\mathrel{\raise.3ex\hbox{$>$\kern-.75em\lower1ex\hbox{$\sim$}}}}
\def\ie{{\it i.e.\ }}
\def\eg{{\it e.g.\ }}

\begin{document}

\tightenlines

\topmargin = -1.0cm
\overfullrule 0pt

\preprint{\vbox{\hbox{IFUSP-DFN/00-053}
\hbox{hep-ph/0007270}
\hbox{Revised July, 2000}
}}
\title{
Three Flavor Long-wavelength Vacuum Oscillation Solution 
to the Solar Neutrino Problem}  

\author{ 
A.\ M.\ Gago$^{1,2}$~\thanks{E-mail: agago@charme.if.usp.br},
H.\ Nunokawa$^{3}$~\thanks{E-mail: nunokawa@ifi.unicamp.br}, and
R.\ Zukanovich Funchal$^{1}$~\thanks{E-mail: zukanov@charme.if.usp.br} }

\address{\sl 
$^1$ Instituto de F\'{\i}sica, Universidade de S\~ao Paulo \\
C.\ P.\ 66.318, 05315-970 S\~ao Paulo, Brazil\\
$^2$  Secci\'{o}n F\'{\i}sica, Departamento de Ciencias, Pontificia 
Universidad Cat\'olica del Per\'u \\
Apartado 1761, Lima, Per\'u   \\
$^3$ Instituto de F\' {\i}sica Gleb Wataghin, 
Universidade Estadual de Campinas, UNICAMP\\    
13083-970 -- Campinas, Brazil.}

\maketitle
\vspace{-0.4cm}
\hfuzz=25pt
\begin{abstract}
We investigate the current status of the long-wavelength vacuum 
oscillation solution to the solar neutrino problem and 
to what extent the presence of a third neutrino can 
affect and modify it. Assuming that the smaller mass squared 
difference that can induce such oscillations, $\Delta m^2_{12}$,
 is in the range $10^{-11}-10^{-8}$  eV$^2$ and the larger one, 
$\Delta m^2_{23}$, in the range relevant to atmospheric neutrino 
observations, we analyze the most recent solar neutrino data coming 
from Homestake, SAGE, GALLEX, GNO and Super-Kamiokande experiments in the 
context of three neutrino generations. 
We include in our vacuum oscillation analysis 
the MSW effect in the Sun, which is relevant for some of the parameter 
space scrutinized. We have also performed, as an extreme exercise, the 
fit without Homestake data. 
While we found that the MSW effect basically does not affect the best 
fitted parameters, it significantly modifies the allowed parameter space 
for $\Delta m^2_{12}$ larger than $\sim 3 \times 10^{-10}$ eV$^2$, in good 
agreement with the result obtained by A.\ Friedland in the case of two 
generations.
Although the presence of a third neutrino does not essentially improve the 
quality of the fit, the  solar neutrino data alone can 
give an upper bound on $\theta_{13}$, which is constrained to be less 
than $\sim 60^\circ$ at 95 \% C.L. 

\end{abstract}
\pacs{PACS numbers:}

\newpage

\section{Introduction}
\label{sec:intro}
The solar neutrino problem (SNP)~\cite{JNB} seems now to be established 
as a definite signal of non-standard neutrino properties 
by the four first generation solar neutrino experiments, 
Homestake~\cite{homestake}, Kamiokande~\cite{kamsolar}, GALLEX~\cite{gallex}, 
SAGE~\cite{sage} and by the new generation higher statistics 
solar neutrino experiments, Super-Kamiokande~\cite{sk00} and GNO~\cite{gno},
 which have strengthened the existence of the SNP. 
Astrophysical explanations of the SNP, which requires significant deviation 
from the standard solar model (SSM)~\cite{BP98,otherSSM}, are 
highly excluded with  the current solar neutrino data~\cite{mina98}. 
Moreover, there is an excellent agreement between 
the sound velocity predicted by the SSM and that obtained from the 
recent helioseismological observations~\cite{BPBC}, which supports 
the rigidness of the SSM. 

Albeit electron neutrinos, produced in the Sun, most certainly 
vanish on their way to the Earth, which is the dynamical origin of the 
process that promotes their disappearance is yet to be completely clarified.  
It has been discussed that the SNP can be nicely explained by the 
simplest extension of the standard electroweak model which invokes neutrino 
mass and flavor mixing~\cite{MNS}. The most plausible solutions in this 
context are provided either by the matter enhanced resonant neutrino 
conversion, the MSW effect~\cite{MSW}, or by the vacuum 
oscillation~\cite{pontecorvo} with a typical wavelength as long as the 
Sun-Earth distance~\cite{justso}.  
Recent detailed analyses and discussions of the solar neutrino data 
based on these mechanisms, \ie  the MSW and long-wavelength vacuum 
oscillation (LVO) solutions, in the context of two neutrino generations,  
can be found, in Refs.\ \cite{BKS,GHPV} and 
Refs.\ \cite{BKS,BFL}, respectively. 
Earlier detailed analyses, prior to the Super-Kamiokande experimental 
results, can be found in Refs.\ \cite{msw_older,vo_older}. 
Discussions on other possibilities to explain the SNP by invoking more exotic 
properties such as neutrino magnetic moment, flavor changing interactions or 
violation of the equivalence principle, can be found, for example, 
in Ref.~\cite{Others}. 

Although it seems at present that pure $\nu_\mu \to \nu_\tau$ oscillations 
in vacuum are quite enough to account for the atmospheric 
neutrino anomaly~\cite{SKatm,atmValencia}, 
the possibility of having contributions from non-negligible 
$\nu_\mu \to \nu_e$ oscillations is still not 
discarded~\cite{fogliatm3nu,yasuda3nu}, 
even after taking into account the constraints coming from the CHOOZ reactor 
experiment~\cite{chooz}. Moreover, the existence of at least three 
neutrino flavors, which is one of the most impressive results from the 
LEP experiments~\cite{PDG98,lep}, makes it unavoidable to try to 
understand neutrino oscillations in a full three generation scenario.

In the framework of three generations of massive neutrinos subjected to  
flavor mixing, one has to deal, in general, with 
six variables to study the neutrino oscillation phenomena, 
namely, three mixing angles, $\theta_{12}$, $\theta_{23}$, $\theta_{13}$ 
and one CP violating phase $\delta$, 
which relate mass and flavor eigenstates,
and two (independent) mass squared differences, 
which can be chosen as $\Delta m^2_{12} \equiv m^2_2- m^2_1$ and  
$\Delta m^2_{23} \equiv m^2_3- m^2_2$. 
If we assume that the smaller mass squared difference,  
defined as $\Delta m^2_{12}$, is in the range relevant 
to solar neutrino oscillation, 
\ie $\Delta m^2_{12}\sim 10^{-11}-10^{-8}$ eV$^2$ in the case of the LVO 
solution or $\Delta m^2_{12}\sim 10^{-8}-10^{-5}$ eV$^2$ in the case of 
the MSW solution, and the larger one, $\Delta m^2_{23}$, is in the range 
relevant to the atmospheric neutrino observations,  
\ie $\Delta m^2_{23}\sim 10^{-3}-10^{-2}$ eV$^2$, 
only three of these variables become significant in practice to the 
solar neutrino investigation: $\theta_{12}$, $\theta_{13}$ 
and $\Delta m^2_{12}$ \cite{Lim,Fogli3nu,KP3nu}.

Under such assumption, a detailed analysis of the three flavor LVO 
solution to the SNP was performed 
in Ref.~\cite{barger} in which the best fit was obtained 
when $\theta_{13}$ is zero, this angle being constrained to 
be less than $\sim 30^\circ$ at 95 \% C.L. by the solar neutrino data alone. 
The authors of Ref.~\cite{barger} have in addition taken into account 
the atmospheric neutrino results, and found that the combined three 
generation fit does not lead to an allowed region appreciably different 
from the one obtained by performing two separate 
effective two-neutrino fits to the solar and atmospheric neutrino data. 
See, for instance, Ref.~\cite{rossi} for earlier discussions 
on the three flavor LVO solution to the SNP.  

On the other hand, a detailed three flavor analysis of   
the MSW solution to the SNP was performed in 
Ref.~\cite{3gmsw1} and not long ago its updated 
version has been reported in Ref.~\cite{3gmsw2}. 
The best fit obtained  by this study permitted to 
constrain $\theta_{13} \lsim 55^\circ-60^\circ$ at 95 \% 
C.L. by the solar neutrino data on its own, without taking into 
account the atmospheric neutrino observations~\cite{SKatm} or the 
CHOOZ result~\cite{chooz}.
More recently, a complete analysis considering four neutrinos, three active 
and one sterile, was performed in the context of both MSW as well as LVO 
oscillation solutions~\cite{4msw_vo}, and bounds on mixing angles 
were again obtained solely by the solar neutrino data. 

In this same spirit we re-examine and update in this paper the 
long-wavelength vacuum oscillation solution to the SNP in the 
two generation as well as in the three generation frameworks. 
In this study, we do not take into account the constraints 
on $\theta_{13}$ coming from the Super-Kamiokande atmospheric 
neutrino observations~\cite{SKatm} nor by the 
CHOOZ reactor experiment~\cite{chooz}, 
whose combination gives an upper bound on $\theta_{13}$ corresponding to
$\sin^2 \theta_{13} \lsim $ few \% \cite{fogliatm3nu,yasuda3nu}. 
The main idea of this paper is to constrain 
$\theta_{13}$ by LVO and the solar neutrino data alone, as 
it has already been done for the atmospheric neutrino 
\cite{fogliatm3nu,yasuda3nu} as well as for 
the MSW solar neutrino solutions~\cite{3gmsw2}. 

Moreover, we investigate the effect of taking the $^8$B neutrino flux 
as known only up to a normalization factor, $f_{\text{B}}$, 
renormalizing, in this case, the $^7$Be flux accordingly, 
assuming the solar temperature power-law~\cite{bu}.  
Finally, we also perform an extreme check analysis by neglecting 
completely the solar neutrino rate measured by the Homestake experiment, 
as considered, for instance, in Ref.~\cite{GFM}, 
on the account of it being the only radiochemical experiment which has not
been calibrated with a radioactive source.

There are new ingredients we include in this work, 
which were not considered in previous analyses of 
the three flavor LVO solution. 
We extend the analysis to mass squared differences up 
to $10^{-8}$ eV$^2$ in order to cover the whole relevant 
LVO parameter space. 
Above this value the oscillation probability of solar neutrinos essentially 
does not depend on the distance between the Sun and the Earth as 
discussed in Ref.\ \cite{GFMBe}, and we do not regard such case as a 
LVO solution to the SNP. 
Thereby, we include in our estimations the MSW effect in 
the Sun in the context of the LVO solution, whose importance was 
first pointed out in Ref.\ \cite{Pantaleone} for 
$\Delta m_{12}^2/E$ larger than $3\times 10^{-10}$ eV$^2/$MeV.
The relevance of the MSW effect in determining the LVO allowed 
parameter space has been recently discussed in Ref~\cite{friedland}. 
Due to the presence of matter effect, we cover the whole relevant
range of mixing angle 
$0 \le \theta_{12} \le \pi/2$~\cite{friedland} in contrast to 
the range usually considered 
($0 \le \theta_{12} \le \pi/4$) for the LVO solution. 

We have used in our analysis the most recent solar neutrino data
from the five solar neutrino experiments\footnote
{For the sake of simplicity, we have not included the data from 
the Kamiokande experiment~\cite{kamsolar}, 
which can be justified by the fact that its result is consistent 
with that of SK and its errors are much larger than 
that of the SK experiment.}
: the total rates measured by the Homestake chlorine (Cl) 
detector~\cite{homestake}, by the GALLEX~\cite{gallex}, GNO~\cite{gno} and 
SAGE~\cite{sage} gallium (Ga) detectors, which are summarized 
in Table  \ref{tab1}, as well as all the observations 
performed by the high-statistics Super-Kamiokande (SK) water Cherenkov 
detector (total rate, energy spectrum and zenith angle dependence)~\cite{sk00}.
The pioneering Homestake experiment which has its energy threshold 
at $E_\nu = 0.814$ MeV~\cite{JNB} is mostly sensitive to 
$^8$B neutrinos ($\sim 76$ \% of the total contributions), 
and also can detect 
$^7$Be neutrinos ($\sim 15$ \%). 
The gallium detectors have their threshold 
at $E_\nu = 0.233$ MeV~\cite{JNB} and are 
sensitive to $pp$ ($\sim 54$ \%),  
$^7$Be ($\sim 27$ \%), and also to 
$^8$B neutrinos ($\sim 10$ \%). 
Super-Kamiokande which has its energy threshold at $E_e = 5.5$ MeV, where $E_e$
is the total energy of the recoil electron, is only sensitive to 
$^8$B neutrinos\footnote{
In this work, we do not take into account the possible 
enhancement of $hep$ neutrinos~\cite{hep} in  
the SK spectrum analysis since the current SK data 
do not seem to conclude on the necessity of 
such enhancement~\cite{sk00}.}.

This paper is organized as follows. 
In Sec.\ \ref{sec:gfa} we present the oscillation formalism
we will use to pave the way to the two and three neutrino generation 
analyses of the data. In Sec.\ \ref{sec:ana} we present and discuss the 
general features of our data analysis. 
In Sec.\ \ref{sec:2g} we revise and update the two-generation LVO 
solution also including  the MSW effect in the Sun. 
In Sec.\ \ref{sec:3g} we extend the analysis to three-generation 
neutrino oscillation taking into account the MSW effect. 
In Sec.\ \ref{sec:season} we examine the seasonal effect that can be 
expected at SK, Borexino and KamLAND experiments for the best fitted 
parameters of the LVO solutions we found.
In Sec.\ \ref{sec:ncl} we discuss the LVO solution with two and three 
neutrino flavors disregarding the chlorine data.
Finally we draw our conclusions in Sec.\ \ref{sec:conc}.

\section{Oscillation Formalism}
\label{sec:gfa}

In this section we describe the framework we will use
in the scope of this paper. 

\subsection{Two Flavor Case}
\label{subsec:two}
For the two generation case in which $\nu_e$ is mixed only with $\nu_\mu$ 
(the same argument holds also for the case of  
$\nu_e-\nu_\tau$ mixing), the evolution 
equation of neutrino traveling through matter can be written as  

\begin{equation}
i\frac{d}{dx} 
\left[
\begin{array}{c}
\nu_e \\ \nu_{\mu}
\end{array}
\right] 
=
\left[
\begin{array}{cc}
V_e(x) & \frac{1}{2}\Delta_{12}    \sin2\theta_{12}  \\
\frac{1}{2} \Delta_{12} \sin2\theta_{12}  & \Delta_{12} \cos2\theta_{12}
\end{array}
\right]
\left[
\begin{array}{c}
\nu_e \\ \nu_{\mu}
\end{array}
\right],
\label{eq:evolution0}
\end{equation}
where $\Delta_{12} \equiv \Delta m_{12}^2/2E$, with the mass squared 
difference of the two neutrino mass eigenstates 
$\Delta m_{12}^2=m_2^2-m_1^2$,  $E$ is the neutrino energy, 
$\theta_{12}$ is the mixing angle,
\be
V_e(x) \equiv \sqrt{2} G_F N_e(x) \simeq 7.6 \times 10^{-14} \times 
\left [ \frac{N_e}{1\ \text{mol/cc}}\right]\ \text{eV}^2,
\label{eq:potential}
\ee
is the matter potential 
for $\nu_e$ with $G_F$ and $N_e$ being the Fermi constant and 
the electron number density, respectively.

This equation, in the case of vacuum ($V_e=0$), simplifies to lead to 
the well known formula for the $\nu_e$ survival probability, 
\begin{eqnarray}
P_{\text{2g,vac}}(\nu_e\to\nu_e) 
&&= 1 - \sin^2 2 \theta_{12}
\sin^2 \left(\frac{\Delta_{12}}{2} L\right) \nonumber \\
&&= 1 - \sin^2 2 \theta_{12}
\sin^2 \left( 1.27 \left[\frac{\Delta m_{12}^2}{\mbox{eV}^2}\right]
\left[\frac{\mbox{MeV}}{E}\right]
\left[\frac{L}{\mbox{m}}\right]
\right),
\label{eq:p2nu_vac}
\end{eqnarray}
where $L$ is  the distance traveled by the neutrino.
It is clear that the probability $P_{\text{2g,vac}}$ 
is invariant under the transformation 
$\Delta m^2_{12} \to -\Delta m^2_{12}$ as well as 
$\theta_{12} \rightarrow \frac{\pi}{2} - \theta_{12} $ 
and therefore, in vacuum it is sufficient to assume 
$\Delta m^2_{12} > 0$ and $0 \leq \theta_{12} \leq \frac{\pi}{4}$ 
to account for all physical situations. 

{}From Eq.~(\ref{eq:p2nu_vac}) we can estimate the 
vacuum oscillation length as, 
\begin{eqnarray}
L_{\text{osc}} &= &\frac{4 \pi E}{\Delta m^2_{12}}
\simeq 2.47 \times 10^{10} 
\left[ \frac{ E}{1\ \text{MeV} } \right] 
\left[ \frac{ 10^{-10} \text{eV}^2} {\Delta m^2_{12} } 
\right] \text{m} \nonumber \\
& \simeq & 0.165 \, L_0
\left[ \frac{ E}{1\ \text{MeV} } \right] 
\left[ \frac{ 10^{-10} \text{eV}^2} {\Delta m^2_{12} } 
\right], 
\label{eq:L_osc}
\end{eqnarray}
where $L_0 \equiv 1 \text{ AU} \simeq 1.496 \times 10^{11}$ m, is 
the astronomical unit, namely,  the mean Earth-Sun distance. 

If matter is present, the physics described by the 
neutrino evolution given in Eq.\ (\ref{eq:evolution0}) 
is not invariant under either transformation 
$\Delta m^2_{12} \to -\Delta m^2_{12}$ or 
$\theta_{12} \rightarrow \frac{\pi}{2} - \theta_{12}$. 
However, we can show that the physical consequence of 
Eq.\ (\ref{eq:evolution0}) is invariant under the 
transformation $\theta_{12} \rightarrow \pi - \theta_{12}$
 as well as $(\Delta m_{12}^2, \theta_{12}) \to  
(-\Delta m_{12}^2, \frac{\pi}{2}-\theta_{12})$. 
Therefore, in order to cover all the possible physically meaningful 
parameter region we can assume~\cite{3gmsw1,3gmsw2}, 
\begin{equation}
\Delta m_{12}^2 >0, \ \ \ 0 \leq \theta_{12} \leq \frac{\pi}{2}. 
\label{eq:range}
\end{equation}
This is the range we consider in this work.

In principle, neutrinos on their way from the interior of the Sun to the 
detector on the Earth can be influenced by the solar as well as by the 
Earth matter potentials. 

Let us initially consider what can happen in 
the Sun. 
In Fig.~\ref{prob}(a) we show the iso-survival probability contours 
of $\nu_e$ at the solar surface as a function of $\Delta m_{12}^2/E$ and 
$\sin^2 \theta_{12}$ for pure vacuum oscillation, ignoring  the matter
effect, assuming that neutrinos are created in the solar center. 
In  Fig.~\ref{prob}(b) we show the same type of contours but 
taking into account the matter effect. The probabilities were 
obtained by numerically integrating Eq.\ (\ref{eq:evolution0}), 
using the electron number density one can find in Ref.\ \cite{bhp}, 
again assuming that neutrinos are created in the center of the Sun. 

{}From these two plots, we immediately see that the probability at 
the solar surface can be significantly different for these two 
cases if $\Delta m_{12}^2/E \ge$  few $\times 10^{-10}$ eV$^2/$MeV. 
Therefore, one can not disregard the matter influence in the $\nu_e$ 
survival probability calculation for values of $\Delta m_{12}^2/E$ in 
this range. For smaller values of $\Delta m_{12}^2/E$ the oscillation 
effect is small and the survival probabilities are $\sim 1$, in both 
cases. 
For $\Delta m_{12}^2/E \sim  {\cal O}(10^{-9})$ eV$^2/$MeV 
or smaller, the r\^ole  of the solar matter is to suppress oscillations, 
as  $\Delta m_{12}^2/E$ increases the MSW resonance effect 
comes into play and strong conversion can occur. 
We note that for values of $\Delta m_{12}^2 \gsim 10^{-9}$ eV$^2$, the 
MSW conversion can be significant for lower energy neutrinos, 
such as $pp$ and $^7$Be ones, but it will be very weak for most of the $^8$B 
neutrinos. 
In fact, we can see from Fig.~\ref{prob} (b), that even for
$\Delta m_{12}^2=10^{-8}$ eV$^2$, the highest value considered in this work, 
if we take $E \sim 10$ MeV, the typical 
$^8$B neutrino energy relevant for SK, the survival 
probability of $\nu_e$ at the solar surface is very close to one. 
This implies that at such high energies, 
neutrinos exit from the Sun essentially 
as $\nu_e$. 

Let us now examine what happens in the Earth.
For the averaged electron number density found in the Earth's mantle 
($N_e \sim 2$ mol/cc) or core ($N_e \sim 5$ mol/cc), 
the matter potential $V_e$ is always much larger than 
$\Delta_{12}$, unless $\Delta m_{12}^2$ is very close to $10^{-8}$ eV$^2$ 
for the lowest energy (\ie $pp$) neutrinos (see Eq.~(\ref{eq:potential})),  
leading to a strong suppression of the effective mixing angle in 
the Earth matter. 
This implies that no appreciable $\nu_{\mu,\tau} \to \nu_e$ 
(or $\nu_e \to \nu_{\mu,\tau}$) re-generation effect can occur 
in the Earth, since this can promote, at the most, a few \% change
in the probability for a very limited range of mixing parameters
($\Delta_{12}\sim 10^{-7}$ eV$^2$/MeV and large $\theta_{12}$).
Hence, in this work we only take into account the MSW effect in 
the Sun and neglect it completely in the Earth.

After accounting for the MSW effect in the Sun, 
the $\nu_e$ survival probability at the Earth given in 
Eq.\ (\ref{eq:p2nu_vac}) is modified as follows \cite{Pantaleone}, 
\begin{eqnarray}
P_{\text{2g}} (\nu_e \rightarrow \nu_e)
&& =  \cos^2\theta_{12} |A_{\nu_1}(R_\odot)|^2 
+ \sin^2\theta_{12} |A_{\nu_2}(R_\odot)|^2 
\nonumber \\
& & 
+ \sin 2\theta_{12} 
|A_{\nu_1}(R_\odot) A_{\nu_2}(R_\odot)|
\cos \left(\Delta_{12} L  + \beta \right), 
\label{materia1}
\end{eqnarray}
where $A_{\nu_1}(R_\odot)$ and $A_{\nu_2}(R_\odot)$ are respectively the 
amplitudes of the neutrino state to be found in the mass 
eigenstate $\nu_1$ and $\nu_2$ at the 
solar surface, $L$ is the 
distance between the solar surface and 
the detection point, 
and $\beta \equiv \mbox{Arg}[A_{\nu_1}(R_\odot)A_{\nu_2}^*(R_\odot)]$
corresponds to the phase developed between $\nu_1$ and 
$\nu_2$ after the neutrinos pass the resonance point 
until they reach the solar surface. 
We remark that for the values of $\Delta m_{12}^2$ we are interested 
in this work, in the region where $\nu_e$'s are created, 
typically $r \lsim 0.3R_\odot$, 
because of the condition $V_e \gg \Delta_{12}$, they 
essentially coincide with the mass eigenstate 
$\nu_1 (\sim \nu_e$), and no oscillation occurs
within this region. Hence the final probability amplitudes 
$A_{\nu_1}(R_\odot)$ and $A_{\nu_2}(R_\odot)$ as well as $\beta$ do 
not depend on the exact production point, 
allowing us to assume that all neutrinos 
are created in the solar center~\cite{Pantaleone}. 

The probability in Eq.\ (\ref{materia1}) can be rewritten 
in terms of the flavor amplitudes as, 
\begin{eqnarray}
&& P_{\text{2g}} (\nu_e \rightarrow \nu_e) \nonumber \\ 
&& =  |A_{\nu_e}(R_\odot)|^2 
\left[ 1 - \sin^22\theta_{12} \sin^2\left(\frac{\Delta_{12}}{2}L\right)\right] 
+ |A_{\nu_{\text{x}}}(R_\odot)|^2
\sin^22\theta_{12} \sin^2\left(\frac{\Delta_{12}}{2}L \right) \nonumber \\
& & 
- \sin 2\theta_{12}  |A_{\nu_e}(R_\odot) A_{\nu_{\text{x}}}(R_\odot)| 
\left[ 2 \cos2 \theta_{12} \sin^2 \left(\frac{\Delta_{12}}{2}L\right) 
\cos \beta'
- \sin\left(\Delta_{12} L\right) \sin \beta' 
\right],
\label{materia2}
\end{eqnarray}
where $A_{\nu_e}(R_\odot)$ and $A_{\nu_{\text{x}}}(R_\odot)$ 
are respectively the probability amplitudes of the neutrino to be found in the 
 state $\nu_e$ and $\nu_{\text{x}}\ (\text{x}=\mu,\tau)$ at the solar surface and 
$\beta' \equiv \mbox{Arg}[A_{\nu_e}(R_\odot)A_{\nu_x}^*(R_\odot)]$, 
is the phase difference developed in the flavor basis corresponding to 
the one we have in Eq.\ (\ref{materia1}) between the mass eigenstates, 
namely, $\beta$. 
We note that the following relation holds, 
\begin{equation}
|A_{\nu_e}(R_\odot)|^2 
=  \cos^2\theta_{12} |A_{\nu_1}(R_\odot)|^2 
+ \sin^2\theta_{12} |A_{\nu_2}(R_\odot)|^2 
+ \sin 2\theta_{12} 
|A_{\nu_1}(R_\odot) A_{\nu_2}(R_\odot)|
\cos \beta. 
\end{equation}

In this work, we first compute the values of $A_{\nu_e}(R_\odot)$ 
and $A_{\nu_{\text{x}}}(R_\odot)$ by numerically integrating the neutrino 
evolution equation in Eq.~(\ref{eq:evolution0}) 
with the solar electron density taken from Ref.\ \cite{bhp}
and  then compute the final probability at the Earth by using the 
expression in Eq.\ (\ref{materia2}). We take into account the effect 
due to  the eccentricity of the Earth orbit around the Sun by taking 
the time average over one year of the probability using the time dependent 
distance $L(t)$,
\begin{equation}
L(t) = L_0 \left[ 1 - \epsilon \cos\left(2\pi \frac{t}{T} \right)
\right] + {\cal O}(\epsilon^2),  
\label{eq:eccentricity}
\end{equation}
where $L_0 = 1$ AU,  
$\epsilon  =  0.0167$ is the orbit eccentricity 
and $T =$ one year. We neglect the correction terms of   
${\cal O}(\epsilon^2)$.

Following Ref.\ \cite{friedland}, we restrict the 
range of $\Delta m^2$ from $10^{-11}$ to $10^{-8}$ eV$^2$.
One can easily estimate from Eqs.~(\ref{eq:p2nu_vac}) and 
(\ref{eq:L_osc}), that for the typical energy of $^8$B neutrinos, 
$\sim 10$ MeV, in order to have appreciable oscillation 
effects for solar neutrinos, the $\Delta m_{12}^2$ value must 
be larger than of the order $\sim 10^{-11}$ eV$^2$. This 
sets our lower limit on $\Delta m_{12}^2$.  

It is known (see, for \eg , Ref.~\cite{friedland}) 
that if $\Delta m_{12}^2$ is larger than $\sim 10^{-8}$ eV$^2$,
the final $\nu_e$ survival probability at the Earth,  
averaged over the neutrino energy, would not practically 
depend on the precise value of the Sun-Earth distance since 
the variation in the probability due to the energy spread 
is large enough, even for $^7$Be neutrinos,  
to average out the vacuum oscillation effect. 
Since we do not regard such a case as a LVO solution, this 
sets our upper limit on $\Delta m_{12}^2$.
Also, to cover values of $\Delta m_{12}^2$ up to 10$^{-8}$ eV$^2$ means to 
examine the effect of the solar matter on the LVO solution to the SNP whose 
relevance was already discussed in Refs.~\cite{Pantaleone,friedland}.  
%

\subsection{Three Flavor Case}
\label{subsec:three}

For the case when we consider three flavor mixing, 
the evolution equation of neutrinos in matter can be 
written as, 
\begin{equation}
i\frac{d}{dx} 
\left[
\begin{array}{c}
\nu_e \\ \nu_\mu \\ \nu_\tau
\end{array}
\right] 
=
\left\{
\left[
\begin{array}{ccc}
V_e(x) & 0 & 0 \\
0 & 0 & 0 \\
0 & 0 & 0
\end{array}
\right]
+ 
U \left[
\begin{array}{ccc}
0 & 0 & 0 \\
0 & \Delta_{12} & 0 \\
0 & 0 & \Delta_{13} 
\end{array}
\right] U^{\dagger}
\right\}
\times
\left[
\begin{array}{c}
\nu_e \\ \nu_\mu \\ \nu_\tau
\end{array}
\right],
\label{evolution1}
\end{equation}
where $\Delta_{ij} \equiv \frac{\Delta m^2_{ij}}{2E}$ 
and we take the representation of the Maki-Nakagawa-Sakata~\cite{MNS}  
mixing matrix $U$ which is the leptonic analog of the 
Cabbibo-Kobayashi-Maskawa matrix in the quark 
sector~\cite{CKM} as, 
\begin{equation}
U  =  e^{i\lambda_7 \theta_{23}} e^{i \lambda_5\theta_{13}}
e^{i\lambda_2\theta_{12}}. 
\end{equation}
Explicitly
\footnote{Here we neglect the possible CP violating phase since this 
phase will not affect the $\nu_e$ survival probability, 
even if we take into account the next-to-leading order 
corrections in electroweak interactions~\cite{MW}.},
\begin{equation}
U =  \left[
\begin{array}{ccc}
\hskip -0.30cm c_{12}c_{13} \hskip -0.30cm 
& s_{12}c_{13} 
& \hskip -0.30cm   s_{13}\nonumber\\
-s_{12}c_{23}-c_{12}s_{23}s_{13} &
c_{12}c_{23}-s_{12}s_{23}s_{13}
& s_{23}c_{13}\nonumber\\
s_{12}s_{23}-c_{12}c_{23}s_{13}
&-c_{12}s_{23}-s_{12}c_{23}s_{13}
& c_{23}c_{13}\nonumber\\
\end{array}
\right],
\label{eqn:ckmmatrix}
\end{equation}
where $\lambda_i$ are the SU(3) Gell-Mann's matrices 
and $c_{ij} = \cos\theta_{ij}$, $s_{ij} = \sin\theta_{ij}$. 
We assume  $\Delta m^2_{13}$ to be in 
the range $\sim 10^{-3}-10^{-2}$ eV$^2$, which is consistent 
with atmospheric neutrino observations \cite{atmValencia}. 
Because of this assumption, the relation 
$V_e$, $\Delta_{12} \ll \Delta_{23} \sim \Delta_{13}$
holds. 

In this case, $\Delta_{13}$ is the dominant term in the
Hamiltonian matrix and this leads to the decoupling 
of $\nu_3$ from the remaining two states. 
This means that the oscillation which is driven by the $\Delta_{13}$
term can be simply averaged out in the final survival 
probability of electron neutrinos at the Earth, 
$P_{\text{3g}}(\nu_e\to\nu_e)$,  
yielding the relation \cite{Lim,Fogli3nu}, 
\be
P_{\text{3g}}(\nu_e\to\nu_e) 
= \sin^4 \theta_{13} 
+ \cos^4 \theta_{13} \cdot P_{\text{2g}}(\tilde{\nu}_e\to \tilde{\nu}_e), 
\label{eq:p3nu}
\ee
where $P_{\text{2g}}(\tilde{\nu}_e\to \tilde{\nu}_e)$ is defined as 
the survival probability of the following effective two 
generation system~\cite{Lim,KP3nu,Fogli3nu}, 
\begin{equation}
i\frac{d}{dx} 
\left[
\begin{array}{c}
\tilde{\nu}_e \\ \tilde{\nu}_{\mu}
\end{array}
\right] 
=
\left[
\begin{array}{cc}
\cos^2\theta_{13} \cdot V_e(x) &\  \frac{1}{2}\Delta_{12} \sin2\theta_{12}  \\
\frac{1}{2} \Delta_{12} \sin2\theta_{12}  &\  \Delta_{12} \cos2\theta_{12}
\end{array}
\right]
\left[
\begin{array}{c}
\tilde{\nu}_e \\ \tilde{\nu}_{\mu}
\end{array}
\right],
\label{evolution2}
\end{equation}
where $\tilde{\nu_{\alpha}}$ ($\alpha = e,\mu, \tau$) is defined as, 
\begin{equation}
\left[
\begin{array}{c}
\tilde{\nu}_e \\ \tilde{\nu}_\mu \\ \tilde{\nu}_\tau 
\end{array}
\right] 
=
e^{-i\lambda_5\theta_{13}} e^{-i\lambda_7\theta_{23}}
\left[
\begin{array}{c}
{\nu_e} \\ {\nu_\mu} \\ \nu_\tau
\end{array}
\right]. 
\end{equation}

We compute $P_{\text{2g}}(\tilde{\nu}_e\to \tilde{\nu}_e)$ by using the 
expression in Eq.\ (\ref{materia2}) just by replacing the solar electron 
density as $N_e(r) \to \cos^2\theta_{13} \cdot N_e(r)$. 

We note that we have three independent parameters 
$\Delta m^2_{12}$, $\theta_{12}$ and 
$\theta_{13}$ which can be fitted or constrained by 
the experimental data. 
Regarding  $\theta_{13}$, it is clear that the probability in 
Eq.~(\ref{eq:p3nu}) is invariant under 
$\theta_{13} \to \pi - \theta_{13}$ allowing us 
to restrict the range of $\theta_{13}$ to $0 \le \theta_{13} \le \pi/2$.  
As for $\Delta m^2_{12}$ and $\theta_{12}$, 
we again consider the range given in Eq.~(\ref{eq:range}) 
for the case of two generations, since 
the same argument discussed in the previous subsection also holds  
for $P_{\text{2g}}(\tilde{\nu}_e\to \tilde{\nu}_e)$. 
 
%
\section{Data Analysis}
\label{sec:ana}

The general idea is to perform a $\chi^2$ analysis to fit the oscillation 
free parameters ($\theta_{12}$ and $\Delta m^2_{12}$, in the case of 
two generations, and $\theta_{12}$, $\theta_{13}$ and  
$\Delta m^2_{12}$ in the case of three generations) with 
observed experimental data. 
We investigate in addition the effect of having  an extra normalization 
factor $f_{\text{B}}$ for the $^8$B neutrino flux, which is in 
this case assumed to be free. 
In what follows we will give explicit definitions of the 
$\chi^2$ for the solar rates, the SK recoil-electron spectrum, the SK zenith 
angle distribution as well as the combination, 
to be used in our analysis.

\subsection{Calculation of the rates}
\label{subsec:rates}

We calculate, for the two (three) generation framework, 
our theoretical predictions for the measured 
solar neutrino rates as a function of the two (three) mixing  parameters 
for the gallium and chlorine experiments by folding the neutrino 
oscillation probability $P_{\text{2g}}$ ($P_{\text{3g}}$) 
with interaction cross section and the six solar neutrino fluxes  
corresponding to each reaction, $pp$, $pep$, $^7$Be, $^8$B, $^{13}$N 
and  $^{15}$O as predicted by the standard solar model of Bahcall and 
Pinsonneault~\cite{BP98} (BP98 SSM). 
Other minor sources such as $^{17}$F and $hep$  
are not considered for simplicity. 
For gallium and chlorine experiments, 
we use the neutrino absorption cross sections found in Ref.\ \cite{bhp} 
and for the SK experiment, we use the new calculation of 
$\nu_e,\nu_{\mu,\tau}$ scattering cross section on electrons 
which take into account radiative corrections~\cite{xsec}.

In this case the expected event rate for Ga and Cl detectors that 
should be compared to experimental data is :

\be
R_{\text{x}}^{\text{theo}}=  C_{\text{x}}
\int dE \sigma_{{\text x}}(E) \langle P_{\text{n}}(E) \rangle
\left(f_{\text{B}} \phi_{\text{B}}(E) + f_{\text{Be}} \phi_{\text{Be}}(E) + 
\sum_{\text{j}} \phi_{\text{j}}(E)\right), 
\label{eq:rates1}
\ee 
where ${\text{x=Ga,Cl}}$, ${\text{n=2g,3g}}$, 
$\phi_{\text{B}}(E)$, $\phi_{\text{Be}}(E)$ and  
$\phi_{\text{j}}(E)$ with ${\text{j=}}pp,pep,^{13}{\text{N} \text{ and } ^{15}{\text{O}}}$  
are the neutrino fluxes as a function of the neutrino energy taken from 
Ref.~\cite{BP98}, and $C_{\text{x}}$ is some normalization 
constant determined in such a way that 
$R_{\text{x}}^{\text{theo}}$ is given in Solar Neutrino Unit (SNU).
The numbers $f_{\text{B}}$ and $f_{\text{Be}}$ are respectively the 
normalization constants for the $^8$B and the $^7$Be neutrino fluxes,  
$f_{\text{B}} = f_{\text{Be}}$ = 1 corresponds to the BP98 SSM values.  
Here $\langle P_{\text{n}} \rangle$ means that the  probability has 
been averaged on the neutrino path length to take into account  
the eccentricity of the Earth orbit around the Sun, 
after we have computed the probability as in Eq.~(\ref{materia2}), for 
two generations, or as in  Eq.~(\ref{eq:p3nu}), for three generations,  
with the solar matter effect included. The integral above was performed 
starting at the energy threshold of each experiment.

In our calculations, we have also included the effect due to 
the thermal broadening of about 1 keV of the $^7$Be lines~\cite{JNBbe7} 
when computing the capture rate by performing an extra average over 
the two $^7$Be neutrino energy profiles given in Ref.~\cite{JNBbe7}. 
This is pertinent for large values 
of $\Delta m^2_{12}$ as discussed, for \eg , in 
Refs.\ \cite{gelb-rosen,friedland}.

When $f_{\text{B}}$ is taken to be free we assume the power-law relationship 
between the $^8$B and $^7$Be fluxes, given by 
$\phi^{^7\text{Be}}=(\phi^{^8\text{B}})^{10/24}$~\cite{bu}, in order to 
renormalize the $^7$Be flux by a factor $f_{\text{Be}}$ as 
\be 
f_{\text{Be}}=f_{\text{B}}^{10/24}.
\label{eq:fbe}
\ee

For the SK experiment, the expected solar neutrino event rate, 
normalized by the BP98 SSM prediction, is given by, 

\be
R_{\text{SK}}^{\text{theo}} 
= 
\frac{ \displaystyle 
f_{\text{B}} \int dE_e \int dE_e^\prime\ h(E_e^\prime,E_e) 
\int dE_\nu \phi_{\text{B}}(E_\nu) 
\left( 
\frac{d \sigma_{\nu_e}}{dE_e^\prime}
\langle P_{\text{n}}(E_\nu) \rangle 
+ 
\frac{d \sigma_{\nu_{\text{x}}}}{dE_e^\prime} 
\left[ \ 1-\langle P_{\text{n}}(E_\nu) \rangle \ \right]
\right) 
}
{\displaystyle 
\int dE_e \int dE_e^\prime\ h(E_e^\prime,E_e)  \int dE_\nu 
\phi_{\text{B}}(E_\nu) 
\frac{d \sigma_{\nu_e}}{dE_e^\prime} 
},
\label{eq:rates2}
\ee
where $E_e$ is the observed recoil electron energy, 
$E_e^\prime$ is the true recoil electron energy, 
$h(E_e^\prime,E_e)$ is the electron energy resolution function 
taken from Ref.\ \cite{res} and  $ d \sigma_{\nu_e}/{dE_e^\prime}$,  
$ d \sigma_{\nu_{\text{x}}}/{dE_e^\prime}$ are 
$\nu_e-e $ and $\nu_{\text{x}}-e \, (\text{x}=\mu,\tau)$ scattering cross 
sections taken from Ref.~\cite{xsec}. In 
the integral above we have used 5.5 MeV as the energy 
threshold for SK.

The definition of the $\chi^2_{\text{rates}}$ function to be minimized is 
the same as the one used in Ref.~\cite{chi2} which essentially 
follows the prescription given in Ref.\ \cite{fogli}. Explicitly,  

\be
\chi^2_{\text{rates}} = \sum_{\text{x,y}} 
(R_{\text{x}}^{\text{theo}}-R_{\text{x}}^{\text{obs}})\sigma_{\text{xy}}^{-2}
(R_{\text{y}}^{\text{theo}}-R_{\text{y}}^{\text{obs}}),
\label{eq:chi2r}
\ee
where x(y) runs through four solar neutrino experiments, 
Homestake, SAGE, GALLEX/GNO combined and Super-Kamiokande, 
the theoretical predictions $R^{\text{theo}}_{\text{x(y)}}$
are given in Eqs.\ (\ref{eq:rates1}) and 
(\ref{eq:rates2}) and the experimental values 
$R^{\text{obs}}_{\text{x(y)}}$ 
are given in Table ~\ref{tab1}. 
The error matrix $\sigma_{\text{xy}}$ contains the theoretical and 
experimental uncertainties according to Ref.\ \cite{chi2} where 
theoretical uncertainties are taken from the ones 
given by the BP98 SSM~\cite{BP98}. 

\subsection{Calculation of the SK recoil-electron Spectrum}
\label{subsec:spec}

For the spectrum shape analysis we first define the following quantity, 
\be
S_{\text{i}}^{\text{theo}} 
\equiv 
\frac{\displaystyle
\int_{E_{\text{i}}^{\text{min}}}^{E_{\text{i}}^{\text{max}}} 
dE_e \int dE_e^\prime\ h(E_e^\prime,E_e) \int dE_\nu 
\phi_{\text{B}}(E_\nu) 
\left( 
\frac{d \sigma_{\nu_e}}{dE_e^\prime} 
\langle P_{\text{n}}(E_\nu) \rangle 
+ 
\frac{d \sigma_{\nu_{\text{x}}}}{dE_e^\prime} 
\left[ \ 1-\langle P_{\text{n}}(E_\nu) \rangle \ \right]
\right) 
}
{\displaystyle 
\int_{E_{\text{i}}^{\text{min}}}^{E_{\text{i}}^{\text{max}}} 
dE_e \int dE_e^\prime\ h(E_e^\prime,E_e)  \int dE_\nu 
\phi_{\text{B}}(E_\nu) 
\frac{d \sigma_{\nu_e}}{dE_e^\prime} 
},
\label{eq:SKspec}
\ee
where $E_{\text{i}}^{\text{min}}$ and $E_{\text{i}}^{\text{max}}$ are the 
minimum and maximum observed recoil electron energy in the i-th bin, 
starting at the 5.5 MeV threshold 
for the SK detector. In total we have 18 bins taken at 0.5 MeV 
intervals, except for the last bin which includes all the 
contribution above 14 MeV.

The definition of the $\chi^2_{\text{spec}}$ function to be minimized is 
the same as the one used in Ref.~\cite{GHPV}, that is, 
\be 
\chi^2_{\text{spec}} = \sum_{\text{i,j=1,18}} 
(\alpha S^{\text{theo}}_{\text{i}}-S_{\text{i}}^{\text{SK}})
\sigma_{\text{ij}}^{-2}(\alpha S^{\text{theo}}_{\text{j}}-S_{\text{j}}^{\text{SK}}),
\label{eq:chi2s}
\ee
where $\sigma_{\text {ij}}$ were computed as prescribed in  Ref.\ \cite{GHPV} 
and $S_{\text{i}}^{\text{SK}}$ are the experimental points whose
numerical values are graphically reproduced from Ref.\ \cite{sk00}.
We will not use the new 5 MeV point since the systematic errors of this point 
are still under study by the SK Collaboration.
The extra normalization parameter $\alpha$, which is always taken to be 
free in our analysis, is introduced because here 
we are only interested in fitting the shape of the spectrum. Moreover  
when we combine with the rates, it allows us to avoid double counting the 
information already taken into account in the rate analysis. 

For the SK spectrum analysis, as a good approximation, 
we simply assume that neutrinos exit from the Sun as 
pure $\nu_e$ and use the vacuum probability 
formula in Eq.~(\ref{eq:p2nu_vac}), for two generation, or 
in Eq.~(\ref{eq:p3nu}), for three generation, in this latter case 
removing from our expression the Sun matter effect. 
This is  well justified by the discussion we presented 
in section \ref{subsec:two}. 

\subsection{Calculation of the zenith angle dependence}
\label{subsec:zen}

We define the $\chi^2_{\text{zenith}}$, for the zenith angle dependence, 
as follows,
\be 
\chi^2_{\text{zenith}} = \sum_{\text{i=1,6}} 
\frac{
(\beta Z^{\text{theo}}_{\text{i}}-Z_{\text{i}}^{\text{obs}})^2} 
{\sigma_{Z,\text{i}}^2}, 
\label{eq:chi2z}
\ee
where $Z^{\text{theo}}_{\text{i}}$ is the $\text{i}$-th bin 
theoretical expectation, $Z_{\text{i}}^{\text{obs}}$ is the $\text{i}$-th bin 
observed value, with 5 night bins and 1 day bin which are 
graphically reproduced from Ref.~\cite{sk00} 
and $\beta$ is a free normalization constant introduced for the same reasons 
as in the case of the spectrum analysis. 
As the LVO solution does not imply in practice any zenith angle distortion 
of the data, this will simply provide an extra global constant that will 
increase the final combined $\chi^2_{\text{min}}$. It is important to 
remark that this global constant will only affect the quality of the 
combined fit, bearing no influence on the computed allowed region. 
We obtained $\chi^2_{\text{zenith}}$ = 4.3 with $\beta= 0.47$ for the 
zenith angle distribution.

\subsection{Combined Analysis}
\label{subsec:global}

Finally the combined $\chi^2$ to be minimized is simply defined as the sum 
of the individual ones:
\be 
\chi^2_{\text{comb}} = \chi^2_{\text{rates}} +\chi^2_{\text{spec}} 
+\chi^2_{\text{zenith}}. 
\label{eq:chi2g}
\ee

\subsection{Definitions of the confidence levels 
and $N_{\text{DOF}}$}
\label{subsec:def_cl}

Using the $\chi^2$ functions defined in the previous subsections,  
for the two generation analysis where we have only two mixing parameters, 
($\Delta m^2_{12}$, $\theta_{12}$),  
we use the condition $\chi^2 = \chi_{\text{min}}^2 + \Delta \chi^2$
where $\Delta \chi^2$ = 4.61, 5.99 and 9.21 for 
90, 95 and 99 \% C.L., respectively, 
in order to determine the allowed parameter space. 
On the other hand, for the three generation analysis,
in order to constrain the parameter space spanned by 
three variables ($\Delta m_{12}^2$, $\theta_{12}$, $\theta_{13}$),  
we determine the iso-confidence level surface 
by the condition $\chi^2 = \chi_{\text{min}}^2 + \Delta \chi^2$
where $\Delta \chi^2$ = 6.25, 7.82 and 11.36 for 
90, 95 and 99 \% C.L., respectively, as in Ref.~ \cite{3gmsw2}. 

We note that in this work, we always determine the values of 
$\chi^2_{\text{min}}$ within the range 
$10^{-11} \text{eV}^2 < \Delta m_{12}^2 < 10^{-8} \text{eV}^2$ 
and therefore, do not take into consideration the region 
relevant for MSW solutions ~\cite{BKS,GHPV}. 
This means that in this work, ``{\it a priori}'' we assume that LVO 
is the solution to the solar neutrino problem, so adopting a different
criteria than other authors, see for \eg Ref.~\cite{4msw_vo}, 
to draw the C.L. contours.

Here, we also describe how we compute the number of degrees of 
freedom $N_{\text{DOF}}$ which is relevant to determine the goodness of 
fit (or C.L.). 
We compute $N_{\text{DOF}}$ as follows, 
\begin{equation}
N_{\text{DOF}}  \equiv N_{\text{data}} -  
N_{\text{param}}  - N_{\text{norm}}\ , 
\end{equation}
where $N_{\text{data}}$ is the number of the data points we use in each 
analysis, $N_{\text{param}}$ is the number of mixing parameters to be
constrained, two or three depending on the number of neutrino generations 
considered, and $N_{\text{norm}}$ is the number of extra free normalization 
factors we introduced in the analysis, \ie 
$f_B$ in Eqs.~(\ref{eq:rates1}) and (\ref{eq:rates2}), $\alpha$ 
in Eq.~(\ref{eq:chi2s}) and $\beta$ in Eq.~(\ref{eq:chi2z}). 
%

\section{Results with The Two Generation Scheme}
\label{sec:2g}

In this section we discuss our results for the analysis of the 
solar neutrino data in the context of two generations. 

\subsection{Results with fixed $f_{\text{B}}$}
\label{subsec:2g_fix}

We first present the case when we fix $f_{\text{B}} = 1$, therefore 
$f_{\text{Be}} = 1$. This corresponds to use the BP98 SSM flux values. 
To demonstrate the influence of the solar matter in the computed allowed 
regions,  we will show here our results without and with the MSW effect 
in the Sun. 

In Figs.\ \ref{allowed_2g_fix_nm}(a),(b) and (c) we show the allowed region in 
$\sin^2 \theta_{12}-\Delta m^2_{12}$ parameter space,  
for the rates, spectrum and combined analysis, respectively, without 
taking into account any possible influence of the solar matter, hence 
all the plots are symmetric with respect to $\sin^2 \theta_{12}=0.5$
because of the invariance of the pure vacuum probability under the 
transformation $\theta_{12} \to \pi/2 - \theta_{12}$.

In Figs.\ \ref{allowed_2g_fix}(a) and (b) we show the allowed region in 
$\sin^2 \theta_{12}-\Delta m^2_{12}$ parameter space,  
for the rates and combined analysis, respectively,  
taking into account the MSW effect. 
As expected we see that in this case the shape of the allowed region is not 
symmetric with respect to $\sin^2 \theta_{12}=0.5$, for 
$\Delta m^2_{12} \gsim 3 \times 10^{-10}$ eV$^2$, in good agreement 
with the results obtained in Ref.\ \cite{friedland}.
We also observe that there is a region allowed at 99 \% C.L. for 
$\Delta m^2_{12} > 4 \times 10^{-10}$ eV$^2$ that only appears when 
matter effects are taken into account.

No spectrum analysis with the solar matter effect was 
performed since we have checked that the results are  
essentially the same as the one presented in Fig.\ \ref{allowed_2g_fix_nm}(b). 
This is because of the fact that nearly all neutrinos that are relevant to 
SK exit from the Sun as pure $\nu_e$ and so we can simply use the vacuum 
oscillation formulas as a very good approximation (see the discussions 
in section \ref{subsec:two}).

The values of the $\chi^2_{\text{min}}$,  the best fitted 
parameters,  the number of degrees of freedom 
($N_{\text{DOF}}$), as well as the goodness of the fit, the C.L. in \%,  
for the above discussed situations are shown in 
Table \ref{tab2}. In this table we also include this information for the 
local best fit point we found at $\Delta m^2 = 6.5 \times 10^{-11}$ eV$^2$ 
and $\sin^2 \theta_{12} = 0.2/0.8$ which explain somewhat better the rates 
than the global best fit one. 

The MSW effect does not affect the values of the best fitted parameters 
for the spectrum and rates analysis, given two virtually symmetric 
solutions in $\sin^2\theta_{12}$. This symmetry is broken however 
when we combine rates and spectrum. Since the spectrum data prefers larger 
values of $\Delta m^2_{12}$, the combined best fit point will lay in an 
allowed island slightly deformed by the matter potential.

Comparing the region allowed by the spectrum at 90 \% C.L., given 
in Fig.~\ref{allowed_2g_fix_nm} (b), with the one allowed by the rates 
at the same C.L., given in Fig.~\ref{allowed_2g_fix} (a), we can see 
that in general the region favored by the total rates are disfavored 
by the spectrum information. The SK spectrum data shows practically no 
energy dependent distortion, preferring either larger values of 
$\Delta m^2_{12}$, where the oscillating term gets averaged to one half, 
or smaller  values of $\Delta m^2_{12}$, where the oscillating term vanishes. 
The total rates, however, prefer intermediate values of  $\Delta m^2_{12}$, 
where the energy dependence of the solution provides the proper reduction 
to explain the four data points.
Due to such disagreement between the total rates and the SK spectrum, 
we find it difficult to get a good fit for both at the same time. 
We note that the combined allowed regions are basically formed by  
the parameters which are more consistent with the spectrum data.

In spite of this incompatibility we note that the quality of the 
combined $\chi^2_{\text{min}}/{\text{\small DOF}}=23.9/24$  seems to be 
quite good. 
This can be understood in virtue of the fact that the oscillation 
parameters at this minimum can provide a very good explanation for the 
SK spectrum ($\chi^2_{\text{min}}=10.7$) and zenith angle 
dependence ($\chi^2_{\text{min}}=4.3$), which are the majority of the  
statistical points, even though they give a generally poor explanation 
for the total rates ($\chi^2_{\text{min}}=8.9$). In fact the oscillation 
parameters at the combined minimum correctly predict the $^{71}$Ga rate 
but for Homestake and SK the predicted values are  substantially above 
what is measured by those experiments.

\subsection{Results with arbitrary $f_{\text{B}}$}
\label{subsec:2g_free}

For the case where we consider $f_{\text{B}}$ as a free parameter 
we also have performed the same analysis without and with matter effect 
taken into account. Although the allowed regions for the rates and the 
combined  analysis increased a little bit they are substantially the same 
as the ones shown in Figs.\ \ref{allowed_2g_fix_nm} and  \ref{allowed_2g_fix}
so we do not show them here.

Again, the values of the $\chi^2_{\text{min}}$ as well as of the best fitted 
parameters can be found in Table \ref{tab2}. 
In comparison to the fixed $f_{\text{B}}$ case, the fit for the rates is 
substantially improved and the allowed parameter region became somewhat 
larger. The other qualitative features mentioned 
in the previous subsection remain unchanged.

The combined fit also improved a bit by this extra freedom 
given $\chi^2_{\text{min}}/{\text{\small DOF}}=21.7/23$ 
(see Table \ref{tab2}), which seems to be even better than in the 
previous case.
Allowing the $^{8}$B flux to be free does not affect the fit of 
the SK spectrum and zenith dependence data, but it improves a little 
the fit for the rates ($\chi^2_{\text{min}}=6.7$) since we can  now  
correctly predict the $^{71}$Ga and SK rates but still over estimate 
the Homestake value.

\section{Results with the Three Generation Scheme}
\label{sec:3g}

In this section we discuss our results for the analysis of the 
solar neutrino data in the context of three neutrino generations. 
Here we will always present the results taking into account the solar 
matter effect.  

\subsection{Results with fixed $f_{\text{B}}$}
\label{subsec:3g_fix}

As for the two generation case we first consider $f_{\text{B}}$ 
as fixed to be 1.
In Fig.\ \ref{allowed_3g_rates_fix} we show the regions, in 
$\sin^2 \theta_{12}-\Delta m^2_{12}$ parameter space, allowed by the 
rates for different values of $\sin^2 \theta_{13}$. 
In each plot of Fig.\ \ref{allowed_3g_rates_fix} we are 
presenting a {\it cut} of the allowed parameter space in three generations, 
that is, we show the allowed region in the $\sin^2\theta_{12}-\Delta m^2_{12}$ 
plane for a given value of $\sin^2\theta_{13}$,  which means that 
the $\chi^2_{\text{min}}$ value is common for all the plots shown.
We found that for the rates 
the best fit occurs when $\sin^2\theta_{13} = 0$.  

We note that, as in the case of the two generation analysis, the 
allowed regions that appear in the plots are asymmetric for 
$\Delta m^2_{12} \gsim 3 \times 10^{-10}$ eV$^2$
because of the matter effect in the Sun. 
Moreover, there is a tendency that the matter influence will gradually start 
to be important at smaller values of $\Delta m^2_{12}$ as $\sin^2 \theta_{13}$ 
increases. We can understand this by recalling that the effective potential 
here has been rescaled by $\cos^2 \theta_{13}$ (see Eq.\ (\ref{evolution2})), 
which means that as  $\sin^2 \theta_{13}$ increases, the potential 
decreases and can start to be relatively relevant for lower values 
of $\Delta m^2_{12}$. 

In Fig.\ \ref{allowed_3g_spec} the  regions, in 
$\sin^2 \theta_{12}-\Delta m^2_{12}$ parameter space, allowed by the 
spectrum for different values of $\sin^2 \theta_{13}$ are displayed. 
Since we have used the vacuum oscillation formulas 
in our spectrum calculation, the allowed regions are symmetric 
with respect to $\sin^2 \theta_{12}=0.5$.   
Again, as in the case of the rates analysis, we found 
the best fit at $\sin^2\theta_{13} = 0$.  
We note that although the best fit occurs when $\theta_{13} = 0$, 
the region excluded by the spectrum data 
become smaller as $\theta_{13}$ increases. 
This can be qualitatively understood from the fact that the energy 
dependence of the probability becomes weaker as $\theta_{13}$ 
increases (see Eq.\ (\ref{eq:p3nu})) and consequently, the spectrum become  
flatter, which is consistent with the present SK data. 

In Fig.\ \ref{allowed_3g_comb_fix} we plot the combined allowed 
region for rates and  spectrum, for fixed $f_{\text{B}}=1$.  
The values we have obtained for the best fitted parameters 
and $\chi^2_{\text{min}}$ as well as the estimated C.L. of the fit are 
presented in Table \ref{tab2}. 
We have found that for the combined analysis, the best fit occurs 
when $\sin^2 \theta_{13}=0.12$. Notwithstanding, a small difference 
in $\chi^2$ such as $\chi^2_{\text{min}}(\theta_{13} = 0)
- \chi^2_{\text{min}}(\theta_{13} = 0.12) \simeq 0.6$ clearly bears no  
real statistical significance to the preferred non-zero $\theta_{13}$.

Finally we plot in  Fig.\ \ref{spec_3g_fix} the spectrum 
calculated for the best fitted parameters for the rates, spectrum and 
combined analysis, for fixed values of $\sin^2 \theta_{13}$, as well as 
the SK experimental data points.  
Here some comments are in order. 
One of the features of  the total rate analysis is that several ``local'' 
best fit points, which are rather comparable in terms of $\chi^2$ values,
exist.
For $\theta_{13}=0$ we found the best fit 
at ($\Delta m^2_{12}$, $\sin^2\theta_{12}$) 
= (9.7$\times 10^{-11}$ eV$^2$, 0.35/0.65). 
We have noticed that as we increase the value of 
$\theta_{13}$ these points do not always remain as 
best fit points but always remain as local best. 
Moreover, such two symmetric (with respect to $\theta_{12}=45^\circ$) best 
fit points move smoothly toward the direction of $\theta_{12}=0$ and 
$\theta_{12}=90^\circ$. 
The fitted spectrum curves for the rates,  shown in  Fig.~\ref{spec_3g_fix}, 
are,  strictly speaking, indicating the ``evolution'' of the spectrum 
shape of these ``best fit points'' as $\theta_{13}$ increases.

We see that the best fit parameters for the rates do not produce a spectrum 
shape  which is consistent with the SK data for lower values  of  
$\sin^2 \theta_{13}$, but the combined fit, which is dominated by the 
spectrum weight in the $\chi^2$ is in fairly good agreement with the data.   

As in the two generation case in spite of the incompatibility between 
rates and spectrum the quality of the combined fit where we found 
$\chi^2_{\text{min}}/{\text{\small DOF}}=23.3/23$ is quite good. Here again 
the oscillation parameters at the minimum can provide a very good 
explanation for the SK spectrum ($\chi^2_{\text{min}}=11.1$) and 
zenith angle dependence ($\chi^2_{\text{min}}=4.3$), the majority of 
the statistical points, even though they give a generally poor explanation 
for the total rates ($\chi^2_{\text{min}}=7.9$),
predicting total rates that are consistent with $^{71}$Ga 
but inconsistent with Homestake and SK data at one or two sigma.

\subsection{Results with arbitrary $f_{\text{B}}$}
\label{subsec:3g_free}

The same plot as in Fig.~\ref{allowed_3g_comb_fix} is presented for 
the combined analysis in Fig.\ \ref{allowed_3g_comb_free},  
when we considered $f_{\text{B}}$ to be free. 
We do not show the rates or the spectrum allowed regions in this case, 
since they are virtually the same as the ones shown in 
Figs.~\ref{allowed_3g_rates_fix} and \ref{allowed_3g_spec}, respectively.

We plot in Fig.~\ref{chi2_rates_fb}, $\chi^2-\chi^2_{\text{min}}$ 
for the rates as a function of $f_{\text{B}}$ for 
three values of $\theta_{13}$. 
{}From this plot, we can see that as $\theta_{13}$ becomes 
larger the allowed values of $f_{\text{B}}$ become 
more restricted, eventually, narrower than the
range allowed by the BP98 SSM, $0.58 < f_{\text{B}} < 1.57$ 
at 3 $\sigma$~\cite{BP98}. 
The calculated spectrum for the best fitted parameters for the rates, 
spectrum and combined analysis, for various values of $\sin^2 \theta_{13}$, 
are similar to the case with $f_{\text{B}}$ = 1, except that the best fit 
point for the total rates can predict a less distorted spectrum shape at lower 
values of $\sin^2 \theta_{13}$ than in the previous case. We find 
it unnecessary to show this here.

The combined fit improved a little with respect to the previous case 
given $\chi^2_{\text{min}}/{\text{\small DOF}}=21.7/22$. 
Again allowing the $^{8}$B flux 
to be free does not affect the fit of the SK spectrum and zenith 
dependence data, but it improves slightly the fit for the 
rates.

\subsection{Constraining $\theta_{13}$}
\label{subsec:theta13}

The values of the $\chi^2_{\text{min}}$ as well as of the best fitted 
parameters for all the above cases are shown in Table \ref{tab3}. 
The combined $\chi^2$ value includes $\chi^2_{\text{zenith}}$ as for 
the two generation case.

To illustrate the effect of the presence of the third 
neutrino in the $\chi^2$,  we plot in Fig.\ \ref{chi2min-fixo} the values of 
$\Delta \chi^2 \equiv \chi^2-\chi^2_{\text{min}}$ as a 
function of $\sin^2 \theta_{13}$ for the cases with $f_{\text{B}}$ fixed.  
{}From these plots, we see that there is a general tendency that the 
fit become worse as $\theta_{13}$ increases, although we note that 
there is a local minimum for the fit with the rates and the combined data at 
$\sin^2\theta_{13} \sim 0.67$. 
We also see that $\Delta \chi^2$ for the rates increases more rapidly 
than that for the spectrum, which rises in fact quite smoothly. 
This can be understood as follows. 

As pointed out in Ref.~\cite{rossi}, naively, the fit for 
the rates is expected to become worse as $\theta_{13}$ increases 
since any energy dependence in the probability will become 
weaker (see Eq. (\ref{eq:p3nu})) and thus, in general, the  
larger the value of $\theta_{13}$, the similar will become the suppressions 
for all the solar neutrinos, leading to a gradually stronger inconsistency 
with the observed total rates. 

On the other hand, as we have already mentioned in 
the previous subsection \ref{subsec:3g_fix}, the loss of energy dependence 
will not compromise so much the fit for the spectrum data  
since the observed spectrum shape is consistent with a flat one. 

Finally, from Fig.\ \ref{chi2min-fixo}, we can conclude that the 
solar neutrino data alone give the upper bound  
$\theta_{13} \lsim 63^\circ$ at 95 \% C.L. 
We have also performed the same analysis for arbitrary 
$f_{\text{B}}$ and obtained, in this case, 
a slightly weaker upper bound, $\theta_{13} \lsim 65^\circ$ at 95 \% C.L.
It is interesting to observe that this limit is quite similar to the one 
obtained in the case of a three flavor MSW solution to 
the SNP~\cite{3gmsw2}, although this  MSW analysis was performed with 
different data.

\section{Predictions for Seasonal Variations}
\label{sec:season}

In this section we discuss, in view of our previous presented results, 
possible seasonal variations that could be measured by the current 
Super-Kamiokande detector,  as well as by the future Borexino~\cite{borexino} 
and KamLAND~\cite{kamland} experiments, which will be sensitive to $^7$Be 
neutrinos. 
It is clear that because of the eccentricity of the Earth 
orbit around the Sun, solar neutrino fluxes should vary 
as $1/L^2$ as $L$ varies with time. 
{}From Eq.~(\ref{eq:eccentricity}) we can estimate that 
the flux variation in one year due to this effect is, 
\begin{equation}
\Delta \phi_\nu
\sim \Delta \left(\frac{1}{L^2} \right) 
\sim \frac{4 \epsilon L_0 }{L^3} 
\sim 4 \epsilon \phi_\nu
\sim 0.07 \phi_\nu, 
\end{equation}
which implies that the solar neutrino signals could
vary as much as $\sim 7$ \% in one year. 

We can see that if vacuum oscillation is assumed,  
``anomalous'' time variation can be expected 
because the expressions of the probabilities 
in Eqs.~(\ref{eq:p2nu_vac}),(\ref{materia1}) and (\ref{materia2})
also dependent on the distance $L$
~\cite{pontecorvo,BPW81,justso}. 
Such ``anomalous'' time variation is expected to be more 
prominent for $^7$Be neutrinos~\cite{PP}. 
This is because of the fact that $^7$Be neutrinos are 
mono-energetic at $E = 0.862$ MeV (and also at $E = 0.383$ MeV,  
with much smaller flux) with $\Delta E \sim 1$ keV~\cite{gelb-rosen}, 
and the oscillation probability of such neutrinos 
can be very sensitive to the precise value of $L$, 
in contrast to the other neutrino fluxes such as 
$pp$ and $^8$B where the probability for these neutrinos 
must be averaged out over the neutrino energy. 
Detailed analyses on the seasonal variations for 
$^7$Be neutrinos have been performed in Refs.~\cite{res,KP,GFMBe}.

In Fig.\ \ref{seassk} we plot the seasonal variation that is 
expected at SK for the best fitted parameters (global and local) of 
the two and three generation LVO solutions for $E_e>5.5$ MeV jointly 
with the observed SK data~\cite{sk00}. The global and local best 
fitted parameters  correspond to the entries Combined and Combined (R), 
respectively, of Table \ref{tab2}, for the two generation, and of Table 
\ref{tab3}, for the three generation solution, 
with $f_{\text{B}}=1$ in both cases.
SK presents the data without subtracting the expected ``normal'' seasonal 
variation due to the $1/L^2$ dependence of the flux. 
For this reason we have included this extra effect in our curves in 
Fig.\ \ref{seassk}.
We see that our predictions are currently consistent with the observed 
SK data.

We also have computed the ``anomalous'' seasonal variation for  
the rate of $^7$Be neutrino flux at 0.862 MeV that one can expect 
to be measured by the Borexino~\cite{borexino} and the 
KamLAND~\cite{kamland} experiments using the best fitted 
parameters (global and local) for the LVO solution with two and three flavors.
Here we have subtracted the expected ``normal'' seasonal 
variation due to the $1/L^2$ dependence of the flux. 
We note that the predictions for Borexino and KamLAND we have obtained 
are visually indistinguishable, and therefore, we only plot the 
case of Borexino in Fig.\ \ref{seas_be}. 
In this figure our predictions are normalized to the BP98 SSM
$^7$Be neutrino flux value, \ie  $f_{\text{Be}} = 1$. 
If $f_{\text{Be}}$ is different from unity, our predictions must be 
renormalized accordingly, keeping the same shape. 
As we can see from these plots, the best fit for two or three neutrino 
generations give similar shapes, also large time variation 
is obtained,  which should be observable at both of these experiments.
Such drastic variation is due to the fact that 
the oscillation wavelength for $^7$Be neutrinos for our global 
best fitted values of $\Delta m^2_{12} = 4.6 \times 10^{-10}$ eV$^2$
and $6.6 \times 10^{-10}$ eV$^2$, is about 3-4 \% of the mean Sun-Earth 
distance (see Eq.~(\ref{eq:L_osc})), 
which is comparable to the Sun-Earth distance variation due to the orbit 
eccentricity.  We have taken into account in our calculations the finite 
width of the $^7$Be line~\cite{JNBbe7}.

Such ``anomalous'' seasonal variation can be a clear signature of the 
vacuum oscillation, which does not depend on any detail of the SSM 
nor on unknown experimental systematic errors~\cite{GFMBe}.

\section{Analysis without Chlorine Data}
\label{sec:ncl}

Finally, in this section, following Refs.\cite{GFM}, 
we have further investigated the impact of removing from our analysis the 
chlorine data, since Homestake is the only radiochemical experiment which 
has not been calibrated with a radioactive source. This is certainly an extreme case but could be worthwhile 
to be discussed.

\subsection{Two generation case}
\label{sec:ncl_2g}

In Figs.\ \ref{allowed_2g_ncl}(a) and (b)  we show our results 
for the rates alone and for the combined analysis, respectively, in 
the case of two generations. We observe here that the rates alone allow 
for a broader range of solutions in the $\sin^2 \theta_{12}-\Delta m^2_{12}$ 
plane, compared to Fig.\ \ref{allowed_2g_fix}(a) where the chlorine data 
was included, making it more consistent with the SK spectrum measurement.  
The combined analysis without chlorine data 
reflects this fact by allowing for larger values of $\Delta m^2_{12}$ even 
at 90 \% C.L.  

The values of the $\chi^2_{\text{min}}$ as well as of the best fitted 
parameters here are shown in Tables \ref{tab4} and Table \ref{tab5}. 
Again the combined $\chi^2$ value also includes $\chi^2_{\text{zenith}}$. 
We remark that without chlorine data the quality of the combined 
fit improves since there are significant overlap between 
the allowed parameter region which give good fit to the total 
rates and SK spectrum. We can see this by comparing 
Figs.~\ref{allowed_2g_fix_nm}(b) and \ref{allowed_2g_ncl}(a).  
This is in contrast to the case with chlorine data.   

\subsection{Three generation case}
\label{sec:ncl_3g}

We repeat the same procedure for the three generation case.
In Figs.\ \ref{allowed_3g_comb_ncl} we show our results 
for  the combined  three flavor oscillation analysis, 
for several values of $\sin^2 \theta_{13}$. 
Again we see that if one ignores Homestake's data completely 
a larger range of values of $\Delta m^2_{12}$ and $\sin^2\theta_{12}$ 
can provide a good  explanation  for the measured rates as well as for 
all the solar data combined.

In Fig.\ \ref{chi2min-fixo} we also show the values of 
$\Delta \chi^2 \equiv \chi^2-\chi^2_{\text{min}}$ as a 
function of $\sin^2 \theta_{13}$ for the analysis 
without $^{37}$Cl data. This curve has the same features of 
the one for the case where all the data was considered in the study, and 
allows us to  put an upper limit  on $\theta_{13}$ to be less 
than $\sim 66^\circ$ at 95 \% C.L..

\section{Discussions and Conclusions}
\label{sec:conc}

We have re-examined the status of the SNP in the light of the LVO 
solution, using the most recent solar neutrino data as well as 
the predictions of the BP98 SSM.
We have analyzed the solar neutrino data in the context of two and three 
neutrino flavor oscillations in vacuum, extending previous analyses to covered 
the range $10^{-11} \le \Delta m^2_{12} \le 10^{-8}$ eV$^2$. In doing this  
we have included the MSW effect in the Sun, which plays a relevant 
r\^ole in the calculations when $\Delta m^2_{12}/E \ge $ 
few $\times 10^{-10}$ eV$^2/$MeV. 
When we have considered three neutrino generations we assumed a mass 
hierarchy such that $\Delta m^2_{13}$ should be large enough to 
be relevant for the atmospheric neutrino problem.

We have found that the MSW effect in the Sun significantly modify 
the allowed parameter space for $\Delta m^2_{12}$ larger than 
$\sim 3 \times 10^{-10}$ eV$^2$ in agreement with the result 
obtained in Ref.~\cite{friedland}. 
We, however, found that values of the best fitted parameters 
are not practically affected by the presence of the solar matter 
effect (see Table II). 

In the two generation LVO solution we found that the rates prefer 
lower values of $\Delta m^2$, \ie $\Delta m^2_{12} \lsim 3 \times 10^{-10}$ 
eV$^2$ at 90 \% C.L., while the SK electron recoil spectrum data 
excludes such favored parameter region and 
prefers $\Delta m^2_{12} \gsim 6 \times 10^{-10}$ eV$^2$.
In the combined analysis the weight of the spectrum prevails. 

The disagreement between the best fitted rates and the spectrum 
predictions for the LVO solution in the case of two generation is, 
nevertheless, less striking when three neutrinos are considered. 
When $\theta_{13}$ is small, the picture described above 
applies but as $\theta_{13}$ increase the region favored 
by the total rates are less excluded by the spectrum data 
compared to the two generation case 
(see for \eg , Fig. \ref{allowed_3g_rates_fix} and 
Fig. \ref{allowed_3g_spec} for the case
 $\sin^2 \theta_{13} = 0.15$). When $f_{\text{B}}$ is arbitrary 
there is even better agreement between spectrum and rates, the regions 
allowed by them already superpose for $\sin^2 \theta_{13} = 0$. 

Judging by the values of the $\chi^2_{\text{min}}$ we have obtained in our 
combined analyses one could conclude that the LVO solution is 
in quite good shape. We point out, however, that this fact 
should be understood as follows: LVO can provide a very good 
explanation for the spectrum and zenith angle dependence in SK
as well as for the total rate of the $^{71}$Ga experiments, 
these are the  majority of the statistical points used in our analyses, 
but provides a poor explanation for the total rates at Homestake and SK.
The SK rate can also be well explained if one allows for 
$f_{\text{B}} \sim 0.74$.

We found that the fit to the total rates as well as SK spectrum 
are not essentially improved due to the possibility of oscillating  
into a third neutrino (see Fig. \ref{chi2min-fixo}). In particular, 
we obtained the upper bound on $\theta_{13}$ to be less 
than $\sim 63^\circ$ at 95 \% C.L. by the solar neutrino data alone. 
We should remark that this bound is substantially weaker than the bound 
obtained by combining the Super-Kamiokande atmospheric neutrino observations 
with the CHOOZ reactor experiment limit~\cite{fogliatm3nu,yasuda3nu}. 
Hopefully future neutrino oscillation experiments at neutrino 
factories will be able to give a much more precise information 
on $\theta_{13}$~\cite{nufactory}.

\vglue 1.0cm 
\noindent 
Note added: While we were completing this work, we became aware  
of two similar works that appeared in hep-ph~\cite{LP,FLMP}. 

\acknowledgments 
We would like to thank Pedro C.\ de Holanda for useful discussions. 
This work was supported by Funda\c{c}\~ao de Amparo \`a Pesquisa
do Estado de S\~ao Paulo (FAPESP) and by Conselho Nacional de e 
Ci\^encia e Tecnologia (CNPq).



%
%
\vglue -2cm 
\begin{table}
\caption{
Solar neutrino rates observed by Homestake, SAGE and GALLEX/GNO combined as 
well as theoretical predictions for  the the Standard Solar Model by 
Bahcall and Pinsonneault\protect \cite{BP98}.
For SK we show the ratio of the observed flux over the prediction 
of the BP98 SSM. 
}
\begin{center}
\begin{tabular}{ccccc}
Experiment & Observed Rate & Ref. &BP98 SSM Predictions~\cite{BP98} & Units \\
\hline
Homestake & 2.56 $\pm$ 0.23 & \cite{homestake} & $7.7^{+1.2}_{-1.0}$ &SNU \\
SAGE & 75.4 $\pm$ 7.6 & \cite{sage} & $129^{+8}_{-9}$ &SNU \\
GALLEX/GNO & 74.1 $\pm$ 6.8 & \cite{gallex,gno} & $129^{+8}_{-9}$ &SNU \\
Super-Kamiokande & 0.465 $\pm$ 0.015 & \cite{sk00} &  $1.00^{+0.19}_{-0.14}$& 
$5.15\times 10^6$ cm$^{-2}$s$^{-1}$\\
\end{tabular}
\label{tab1}
\end{center}
\end{table}
%
%
\vglue 0.2cm
\begin{table}[h]
\caption[Tab]{
The best fitted parameters and $\chi^2_{\text{min}}$ 
as well as C.L. (in \%) for the 2 generation LVO  solution to 
the SNP. The number of degrees of freedom ($N_{\text{DOF}}$) 
are also indicated. The matter effect was taken into account 
unless it is indicated in parentheses. Note that for 
spectrum analysis matter effect was neglected, as it is a good 
approximation (see section \ref{subsec:two}, \ref{subsec:spec}).  
We also present the values of the local best fit  
which explain better the total rates in the line 
indicated as Combined (R). 
}
\vglue -0.4cm
\begin{center}
\begin{tabular}{ccccccc}
Case  & $\Delta m^2_{12} \times 10^{10}$ eV$^2$ 
& $\sin^2 \theta_{12}$ & $f_{\text{B}}$ 
&$\chi^2_{\text{min}}$ & $N_{\text{DOF}}$ & C.L. (\%) \\
\\ \hline
Rates  & 0.97  & 0.35/0.65 & 1 (fixed) & 0.25 & 2 & 88.2   \\ 
Spectrum   & 6.5   & 0.38/0.62  & --  & 10.5 & 15 & 78.7   \\ 
Combined (w/o matter)   & 6.6 & 0.35/0.65  &  1 (fixed) & 25.1  & 24 & 40.0 \\ 
Combined    & 6.6 & 0.64  &  1 (fixed) & 23.9 & 24  & 46.7  \\ 
Combined (R) & 0.65 & 0.2/0.8  &  1 (fixed) & 28.0 & 24  & 26.0  \\ 
\hline
Rates  & 0.88  & 0.34/0.66 & 1.21  & 0.01 & 1  & 91.6 \\ 
Combined (w/o matter)  & 6.6 & 0.31/0.69  &  0.72 & 22.7  & 23 & 47.8 \\
Combined    & 6.6 & 0.66  &  0.74 & 21.7 & 23  &  53.8 \\
Combined (R)   & 0.65 & 0.2/0.8  &  0.86 & 27.5 & 23  & 23.5  \\ 
\end{tabular}
\end{center}
\label{tab2}
\vglue -0.8cm
\end{table}

%
%

\begin{table}[h]
\caption[Tab]{The best fitted parameters and $\chi^2_{\text{min}}$ for the 
3 generation LVO solution to the SNP. 
}
\vglue -0.4cm
\begin{center}
\begin{tabular}{cccccccc}
Case  & $\sin^2 \theta_{13}$ & $\Delta m^2_{12} \times 10^{10}$ eV$^2$ 
& $\sin^2 \theta_{12}$ &
$f_{\text{B}}$ &$\chi^2_{\text{min}}$ & $N_{\text{DOF}}$ & C.L. (\%) \\ 
\\ \hline
Rates  & 0.0     & 0.97  & 0.35/0.65 & 1 (fixed) & 0.25 & 1 &  61.7 \\ 
Spectrum   &  0.0   & 6.5   & 0.38/0.62  & --  & 10.5 & 14 & 72.5 \\ 
Combined    & 0.12   & 4.6 & 0.75  &  1 (fixed) & 23.3  & 23  & 44.3 \\ 
Combined (R)  & 0.14 & 0.66  &  0.15/0.85 & 1 (fixed) & 26.1 & 23  & 29.6  \\ 
\hline
Rates  & 0.0     & 0.88  & 0.34/0.66 & 1.21  & 0.01 & 0 & --   \\ 
Combined      & 0.0   & 6.6    & 0.66  &  0.74 & 21.7 & 22 & 47.7   \\ 
Combined (R)  & 0.1 & 0.65 & 0.14/0.86 &  0.82  & 25.5   & 22   & 27.4 \\ 
\end{tabular}
\end{center}
\label{tab3}
\vglue -0.8cm
\end{table}

%
%

\begin{table}[h]
\caption[Tab]{The best fitted parameters and $\chi^2_{\text{min}}$ for the 
2 generation LVO solution to the SNP without the chlorine data. }
\vglue -0.4cm
\begin{center}
\begin{tabular}{cccccccc}
Case  & $\Delta m^2_{12} \times 10^{10}$ eV$^2$ & $\sin^2 \theta_{12}$ &
$f_{\text{B}}$ &$\chi^2_{\text{min}}$ & $N_{\text{DOF}}$ & C.L. (\%) \\ 
\\ \hline
Rates  & 0.97  & 0.33/0.67 & 1 (fixed) & 0.01 & 1 &  92.0 \\ 
Combined    & 6.5 & 0.69  &  1 (fixed) & 18.0 & 23 & 75.7   \\ 
\end{tabular}
\end{center}
\label{tab4}
\vglue -0.8cm
\end{table}

%
%

\begin{table}[h]
\caption[Tab]{The best fitted parameters and $\chi^2_{\text{min}}$ for the 
3 generation LVO solution to the SNP without the chlorine data. 
}
\vglue -0.4cm
\begin{center}
\begin{tabular}{cccccccc}
Case  & $\sin^2 \theta_{13}$ & $\Delta m^2_{12} \times 10^{10}$ eV$^2$ 
& $\sin^2 \theta_{12}$ &
$f_{\text{B}}$ &$\chi^2_{\text{min}}$ & $N_{\text{DOF}}$ & C.L. (\%) \\ 
\\ \hline
Rates  & 0.0    & 0.97  & 0.33/0.67 & 1 (fixed) & 0.01 & 0 & --  \\ 
Combined    & 0.1   & 4.6 & 0.76  &  1 (fixed) & 17.4 & 22 & 74.1 \\ 
\end{tabular}
\end{center}
\label{tab5}
\vglue -0.8cm
\end{table}


\newpage


%
\begin{figure}
\caption{ Contours of iso-survival probability of $\nu_e$ 
at the surface of the Sun as a function of $\Delta m^2_{12}/E$ and $\sin^2 \theta_{12}$ (a) 
considering pure vacuum oscillation, and (b) 
when matter is taken into account. For the case (b) we 
numerically integrated the evolution Eq.~(\ref{eq:evolution0}) 
assuming that neutrinos are created at the center of the Sun.
}
\label{prob}
\vglue -0.05cm
\end{figure}

%
\begin{figure}
\caption{
Region of $(\sin^2 \theta_{12}, \Delta m^2_{12})$ allowed by (a) the 
total rates from $^{37}$Cl,  $^{71}$Ga and SK experiments,  
(b) the SK spectrum and (c) the combined analysis 
of rates + SK spectrum 
in the long-wavelength vacuum oscillation scenario 
for  2 neutrino flavors. 
In the case of the SK spectrum (b) it is the inner part of the contours 
which is excluded by the data. The best fit points are shown as 
black circles. Solar matter effect is not taken into account. 
}
\label{allowed_2g_fix_nm}
\vglue -0.05cm
\end{figure}

%
\begin{figure}
\caption{
Same as in Figs.~\ref{allowed_2g_fix_nm} (a) and (c) but with 
the solar matter effect. 
}
\label{allowed_2g_fix}
\vglue -0.35cm
\end{figure}

%
\begin{figure}
\caption{
Region of $(\sin^2 \theta_{12}, \Delta m_{12}^2)$ allowed 
by the total rates from $^{37}$Cl,  $^{71}$Ga and SK
for various values of $\sin^2\theta_{13}$ for the 
long-wavelength vacuum oscillation solution to the SNP 
with 3 neutrino flavors. 
}
\label{allowed_3g_rates_fix}
\vglue -0.25cm
\end{figure}
%
\begin{figure}
\caption{
Region of $(\sin^2 \theta_{12}, \Delta m_{12}^2)$ allowed 
by the SK spectrum for various values of $\sin^2\theta_{13}$ 
for the LVO solution to the SNP with 3 neutrino flavors. 
It is the inner part of the contours which is excluded by the data.
}
\label{allowed_3g_spec}
\vglue -0.35cm
\end{figure}


%
%

\begin{figure}
\caption{
Same as in Fig. \ref{allowed_3g_rates_fix} but 
for the combined analysis of total rates + SK spectrum.  
}
\label{allowed_3g_comb_fix}
\vglue -0.35cm
\end{figure}

%
%
\begin{figure}
\caption{
Expected SK spectrum using the 
best fitted values of $(\sin^2 \theta_{12}, \Delta m_{12}^2)$. 
We note that the spectrum curves determined only by the rates
are adjusted in such a way that $\chi^2$ defined in 
Eq.~(\ref{eq:chi2s}) takes minimum values after we determine 
the values of $(\sin^2 \theta_{12}, \Delta m_{12}^2)$ by the 
fit only with the rates. 
}
\label{spec_3g_fix}
\vglue -0.35cm
\end{figure}

%
%
\begin{figure}
\caption{
Same as in Fig. \ref{allowed_3g_comb_fix} but 
with arbitrary $^8$B neutrino flux normalization 
$f_{\text{B}}$. 
}
\label{allowed_3g_comb_free}
\vglue -0.35cm
\end{figure}


%
\begin{figure}
\caption{
$\chi^2-\chi^2_{\text{min}}$ 
is plotted as a function of 
$f_{\text{B}}$ for various values of $\sin^2 \theta_{13}$. 
We also show, by the vertical lines, the range of allowed 
values of $f_{\text{B}}$ at one and three standard deviations 
as  predicted by the BP98 SSM~\protect\cite{BP98}.} 
\label{chi2_rates_fb}
\vglue -0.35cm
\end{figure}

\begin{figure}
\caption{
$\chi^2-\chi^2_{\text{min}}$ as a function of $\sin^2 \theta_{13}$,  
for the case when $f_{\text{B}}$ is fixed to be equal to 1 for the 
rates (R), spectrum (S), combined analysis (R+S) and for the combined 
analysis without $^{37}$Cl data (R+S noCl).}
\label{chi2min-fixo}
\vglue -0.35cm
\end{figure}
%
\begin{figure}
\caption{Expected seasonal variation at SK  for
the best fitted parameters of the two and three generation LVO solutions 
to the SNP. We also show in the plot the experimental data points.
We have not subtracted the effect of the ``normal'' seasonal variation 
(due to $\sim 1/L^2$ dependence) expected in the absence of any 
oscillation. }
\label{seassk}
\vglue -0.5cm
\end{figure}
\begin{figure}[p]
\caption{Same as in Fig. 11 but for Borexino and KamLAND. Here 
we have subtracted the ``normal'' seasonal variation due to the $1/L^2$ 
dependence of the flux.}
\label{seas_be}
\vglue -0.35cm
\end{figure}

%
%

\begin{figure}
\caption{
Region of $(\sin^2 \theta_{12}, \Delta m^2_{12})$ allowed by (a) the 
total rates from $^{71}$Ga and SK,
(b) the combined analysis of rates + SK spectrum 
in the LVO scenario for  2 neutrino flavors. The best fit points are shown as 
a dark circle. Here we have ignored the information from 
 $^{37}$Cl.  
}
\label{allowed_2g_ncl}
\vglue -0.35cm
\end{figure}


%
\begin{figure}
\caption{
Region of $(\sin^2 \theta_{12}, \Delta m_{12}^2)$ allowed 
by the combined analysis of rates + SK spectrum 
for various values of $\sin^2\theta_{13}$ 
for the LVO solution to the SNP 
with 3 neutrino flavors. Here we have ignored the information from 
 $^{37}$Cl.  
}
\label{allowed_3g_comb_ncl}
\vglue -0.35cm
\end{figure}



\end{document}